\documentclass[twocolumn]{aastex7}

\usepackage{booktabs}

\definecolor{pink}{rgb}{1,0.1,.6}

\begin{document}

\title{Two Validated Planets and 179 Candidates From a TESS Faint Star Transit Search in the ULTRASAT Fields}

\author[0009-0009-4693-7783]{Matthew Lastovka}
\affiliation{Department of Astronomy, University of Maryland, College Park, MD 20742, USA}
\affiliation{NASA Goddard Space Flight Center, 8800 Greenbelt Road, Greenbelt, MD 20771, USA}
\email[show]{mlastovk@umd.edu}

\author[0000-0001-7139-2724]{Thomas Barclay}
\affiliation{NASA Goddard Space Flight Center, 8800 Greenbelt Road, Greenbelt, MD 20771, USA}
\email{thomas.barclay@nasa.gov}

\author[0000-0002-4235-6369]{Robert F. Wilson}
\affiliation{Department of Astronomy, University of Maryland, College Park, MD 20742, USA}
\affiliation{NASA Goddard Space Flight Center, 8800 Greenbelt Road, Greenbelt, MD 20771, USA}
\email{robert.f.wilson@nasa.gov}

\author[0000-0002-0388-8004]{Emily A. Gilbert}
\affiliation{Caltech/IPAC-NASA Exoplanet Science Institute, Pasadena, CA 91125, USA}
\email{dremilygilbert@gmail.com}

\begin{abstract}

The TESS mission is designed to survey the entire sky to find planets most amenable to follow-up observations, and transit searches have therefore focused specifically on bright targets. Faint stars ($T_{mag}>13.5$) have not been systematically searched for planets, despite TESS having the sensitivity necessary to do so, therefore representing a promising avenue for increasing TESS's scientific yield. We utilized TESS's sensitivity for faint stars in particular to find more planets for joint optical-UV studies of exoplanet transits. Observing transits in both UV and optical gives insight into exoplanet atmosphere composition and evolution and remains relatively unexplored due to the lack of UV-capable instruments. ULTRASAT is an upcoming UV space telescope that will perform a high cadence survey suitable for observing exoplanet transits. To increase the sample size of planets capable of observations with ULTRASAT, we performed a transit search of faint stars down to $T_{mag}=16$ in the ULTRASAT high cadence survey fields using TESS light curves optimized for faint stars. Our search recovered 262 planet candidates, of which 181 are new to this work. We also statistically validated 2 of those candidates as bona fide planets. Finally, we estimated the ULTRASAT SNR of both new and existing planet candidates, and found that 113 targets will be detectable with ULTRASAT, which would facilitate the largest study of transiting planets in the visible and UV. This work demonstrates the power of TESS to detect exoplanets around faint stars and provides a catalog of well-vetted transiting planet candidates to observe with ULTRASAT.

\end{abstract}

\keywords{Exoplanets (498) --- Space telescopes (1547) --- Transit photometry (1709)}

\section{Introduction}

Exoplanet transits are less well studied in the ultravoilet (UV) compared to optical and infrared wavelengths since the UV is a notoriously difficult range to observe. Earth's atmosphere absorbs nearly all UV light, necessitating that astrophysical observations at these wavelengths be performed from space. These limitations, in addition to the cost and difficulty of manufacturing UV capable instrumentation, have led to the limited availability of UV data for transiting exoplanet science. Despite the limited number of UV transit observations, they provide valuable insights into exoplanet atmospheres' composition and evolution. Single object follow up (e.g., using the Hubble Space Telescope) can be incredibly useful, but is inherently limited for conducting population-level studies. A wide-field survey has the ability to measure transits from many more sources simultaneously. However, past and current transit searches have only observed at visible and near-infrared wavelengths (e.g. \textit{Kepler}; \citealt{Borucki2010}, TESS; \citealt{Ricker2015}, Roman; \citealt{Spergel2015, Akeson2019}, PLATO; \citealt{Rauer2025}).

UV observations are particularly valuable for understanding exoplanet atmospheric escape. Observations of the \ion{H}{1} Ly$\alpha$ emission line in the FUV (1215.67 \AA) in transit has shown absorption in significant excess of the optical transit. For example, \cite{Ehrenreich2015} observed GJ 436 b in Ly$\alpha$ with the Hubble Space Telescope and observed a 56.3\% transit depth, while the optical transit depth is less than 1\%. These results are inferred to be a result of a large cloud of hydrogen escaping from the planet's atmosphere. To date, a few dozen planets have been observed in Ly$\alpha$, and those observations have provided insights into planetary mass loss rates, the speed and geometry of escaping material, and interactions between the escaping material and the stellar wind \citep[e.g,][]{Bourrier2016, Esquivel2019, Khodachenko2019, Shaikhislamov2021, Rockcliffe2023, Loyd2025}.

Observations of some transiting planets in the near-UV (NUV) have also indicated significantly more absorption than in optical wavelengths. Excess absorption has been observed both in transmission spectroscopy and NUV photometry. Photometric observations have been conducted with observatories such as the Colorado Ultraviolet Transit Experiment (CUTE; \citealt{france2023}) SmallSat, Swift-UVOT \citep{Gehrels2004, Roming2005}, and XMM-Newton. Multiple processes can explain an observed NUV excess, including planetary atmospheric escape, clouds, and hazes \citep[e.g.,][]{Ardila2018, Helling2020, Cubillos2020, Koskinen2022}. For example, \cite{Salz2019} observed the transit of the hot Jupiter WASP-121 b using Swift-UVOT (220-270 nm) in which they observed a NUV transit depth of 2.1\% (larger than the 1.55\% optical transit depth). The observed NUV excess is only slightly smaller than the 2.5\% projected cross-section of the planet's Roche Lobe, indicating that the absorption is due to ongoing atmospheric escape on WASP-121 b. However, atmospheric escape cannot explain the absorption seen on all planets. For example, the hot Jupiter X0-3 b has a 30-70\% deeper transit in the NUV compared to the optical (observed with XMM-Newton; 200-250 nm), which corresponds to a NUV transit depth of 13\% \citep{Cilley2026}. However, this planet has a large Roche radius (projected cross-section of $\sim$65\%), which, when combined with the observed X-ray luminosity of the host star, implies that this planet should have a very small mass loss rate \citep{Cilley2026}. Another cause of the larger transit depth could be absorption from SiO, a precursor to silicate clouds \citep{Lothringer2020}. When silicate clouds evaporate at higher equilibrium temperatures, the SiO will absorb strongly, creating massive excess absorption in the NUV. 

The sample size of planets observed with NUV photometry is very small ($\sim$ 10). Current observatories have limited sensitivity, and can struggle to obtain definitive transit observations. For example, \cite{Corrales2021} observed 5 known hot Jupiters using Swift-UVOT. They reported 1 positive transit detection, 1 marginal detection, and could only place upper limits on the transit depths of the other three. This small sample size makes it difficult to examine trends in the NUV transit depth and how it relates to atmospheric properties, like escape.

The Ultraviolet Transient Astronomy Satellite (ULTRASAT) is a planned space-based, wide-field ultraviolet observatory designed to characterize transient and variable UV sources \citep{Shvartzvald2024}. Key science cases for the mission include shock breakout and early ultraviolet emission from core-collapse supernovae, tidal disruption events, electromagnetic counterparts to gravitational-wave sources, and other rapidly evolving phenomena that are difficult to capture with narrow-field ultraviolet facilities. ULTRASAT aims to provide early detection and continuous light curves of hundreds of supernovae and has fast slewing capabilities to follow up potential binary neutron star mergers identified from gravitational wave detectors, with a field of view large enough to cover the error ellipses from those detections. The mission is led by the Israel Space Agency (ISA) with participation from NASA and international partners \citep{Shvartzvald2024}. ULTRASAT is currently under construction and scheduled for launch in late 2027. It will have a 33 cm aperture, a field of view of 204 deg$^2$ in a single snapshot, and will observe in a single bandpass covering 230-290 nm. This wavelength range covers absorption features from refractory species that can be used to track escape and silicate cloud formation, such as Mg II, Fe II, and SiO. The mission will produce an all-sky map in the near-ultraviolet (NUV), conduct a high-cadence survey covering three fields with $\sim$300-second cadence, and conduct a low-cadence survey covering a larger number of fields at $\sim$1-day cadence.

ULTRASAT's huge field of view and nearly an order of magnitude improvement in photometric precision over current observatories will enable observations of a large number of exoplanet transits. ULTRASAT's high cadence survey will observe a single field for 21 hours per day with continuous 300-second exposures. After observing a given field for 6 months, ULTRASAT will slew to another field and observe that one for 6 months. ULTRASAT will rotate between 3 fields, 2 in the northern hemisphere and one in the southern hemisphere, over the 3-year nominal mission. Figure \ref{fig:fields} shows that the 3 ULTRASAT high-cadence survey fields have considerable overlap with other exoplanet transit observatories, particularly NASA's Transiting Exoplanet Survey Satellite (TESS; \citealt{Ricker2015}) and the PLAnetary Transits and Oscillations of stars mission (PLATO; \citealt{Rauer2025}). At the time of this publication, there are 77 confirmed transiting planets in the ULTRASAT fields, along with 219 planet candidates. To maximize the scientific return of ULTRASAT, we need to increase the sample of known planets in the ULTRASAT fields. Therefore, we search TESS data for additional targets within the ULTRASAT fields. 

The TESS mission, launched in 2018, has identified more than 8000 planet candidates, including more than 900 confirmed planets \citep{Guerrero2021,Christiansen2025}. TESS observes the sky with four wide-field cameras that monitor a 24$^{\circ}$$\times$96$^{\circ}$ sector continuously for roughly 27 days, producing Full Frame Images (FFIs) of the entire field. FFIs were downlinked from the spacecraft and made available at a 30-minute cadence during the Prime Mission, a 10-minute cadence during the First Extended Mission, and are being downlinked at a 200-second cadence during the Second Extended Mission and beyond.

The majority of stars observed by TESS only in FFIs have not been searched for transiting planets because those searches are computationally expensive, and running a search on every star is not feasible for the TESS team. The TESS Science Processing Operations Center (SPOC) prioritizes a smaller number of bright, nearby targets for which TESS has a higher sensitivity to planets and that can be more easily followed up with other ground and space-based observatories \citep{Caldwell2020}. Additionally, the standard vetting pipeline requires manual inspection of candidates, making it infeasible to search for planets around every star. Therefore, many TESS faint stars have not been searched for planets, despite TESS having the photometric precision to detect planets around such stars.

There are several groups that have used FFI data to search for planets around fainter stars. One of the faintest searches is done with the Quick Look Pipeline (QLP), which makes light curves from FFIs for all stars with TESS magnitude $\leq$13.5 and has identified more than 3300 planet candidates \citep{Kunimoto2022c}. In total, transit searches using QLP light curves have identified more than 5300 TESS Objects of Interest (TOIs) to date, representing nearly 70\% of all identified TOIs. However, the cutoff magnitude of 13.5 for this search was set by available computational resources.

To increase the number of planets that will be observable by ULTRASAT, we pushed our TESS transit search to fainter stars. 
Despite the worsened photometric precision for TESS light curves of faint stars, the number of stars increases rapidly as a function of apparent magnitude, meaning we may still detect a large number of planet candidates. In this work, we implemented two strategies to push our transit search to fainter stars. First, we employed {\tt tess-phomo}, a new method for creating light curves from TESS FFIs with improved photometric precision (see Section \ref{sec:tessphomo}), to increase the signal-to-noise ratio of planet transits. Second, we limit our sample size so that we are not forced to compromise the quality of our search to increase efficiency and focus on a modest sample of stars in the ULTRASAT fields.

In this work, we expanded the sample of known transiting planet candidates detectable by ULTRASAT by conducting a faint star transit search using TESS. The structure of the paper is as follows. Section \ref{sec:methods} outlines our transit search and vetting pipeline, and our method for estimating the signal-to-noise ratio (SNR) of transiting planets in ULTRASAT. Section \ref{sec:results} describes the results of our transit search. In Section \ref{sec:discussion}, we discuss our validated planets and other planet candidates of particular note (\S\ref{sec:validated}) and a comparison to the recent results from the T16 project (\S\ref{sec:t16}). Finally, Section \ref{sec:conclusions} presents our conclusions.

\section{Methods} \label{sec:methods}


This section outlines our transit search and vetting pipeline. We discuss our sample selection (\S\ref{subsec:sample}), light curve generation (\S\ref{sec:tessphomo}), transit search pipeline (\S\ref{subsec:tls}), and vetting pipeline (\S\ref{subsec:vetting}). We also describe how we validated our planet candidates with {\tt TRICERATOPS} (\S\ref{subsec:triceratops}) and estimated the ULTRASAT SNR for each candidate (\S\ref{subsec:ultrasat_snr}).

\subsection{Sample Selection}
\label{subsec:sample}

\begin{figure*}
    \centering
    \includegraphics[width=0.9\linewidth]{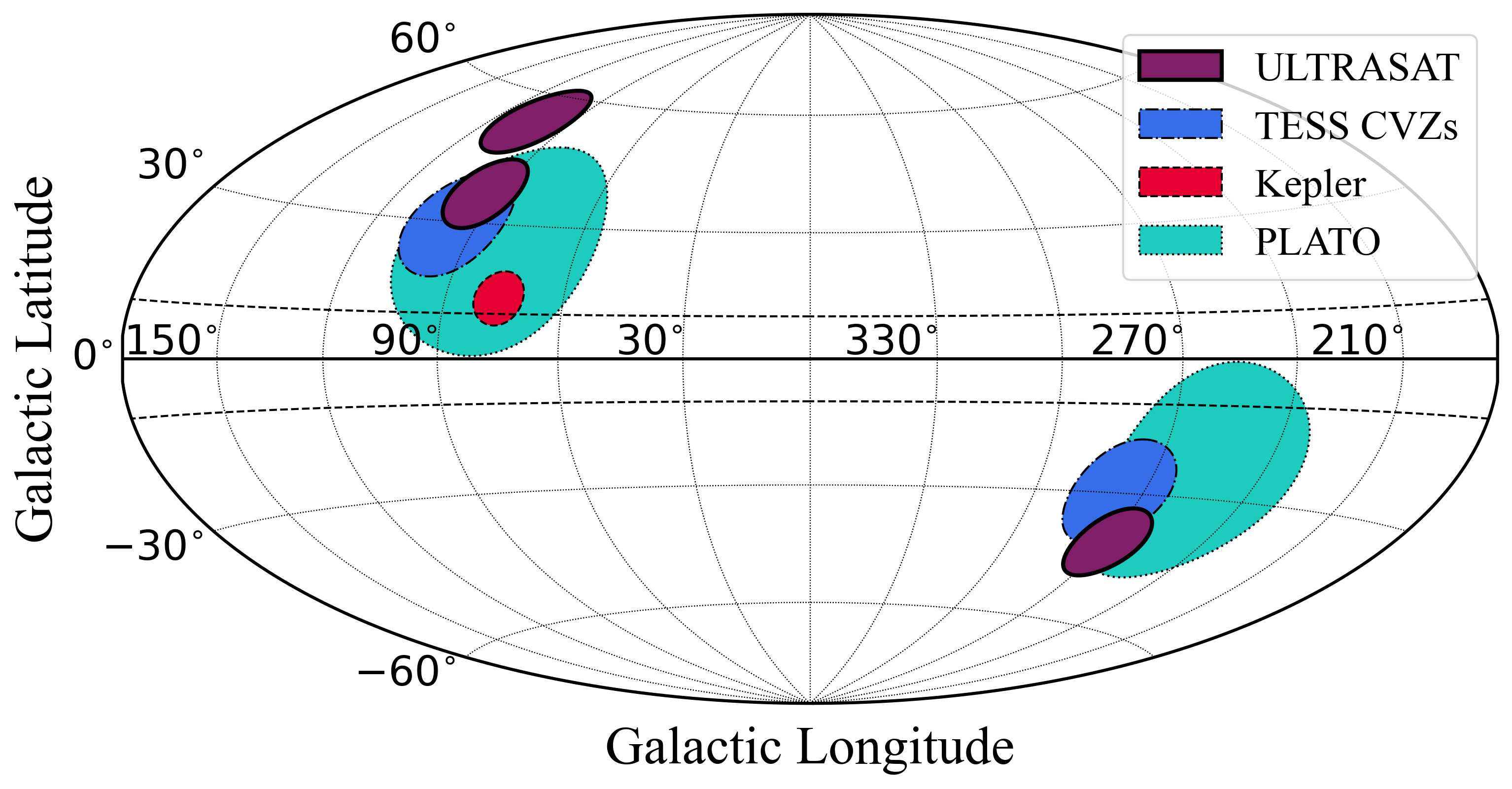}
    \caption{The ULTRASAT fields have considerable overlap with the TESS Continuous Viewing Zones (CVZs) and the PLATO fields. The ULTRASAT fields are shown in purple, the TESS CVZs in blue, the planned PLATO long pointing fields in green, and the Kepler field in red. }
    \label{fig:fields}
\end{figure*}

\begin{table}
\centering
\begin{tabular}{c|cc}
\hline
Field & RA & Dec \\
\hline
S1 & 67$^\circ$ & -59$^\circ$ \\
N2 & 215$^\circ$ & 60$^\circ$ \\
N3 & 254$^\circ$ & 64$^\circ$ \\
\hline
\end{tabular}
\caption{ULTRASAT High Cadence Survey Field Locations. S and N indicate whether the field is located in the southern or northern hemisphere, respectively.}
\label{tab:field_locations}
\end{table}

To create our transit search target list, we queried the TESS Input Catalog (TIC) for stars with TESS magnitude $<$ 16 in the ULTRASAT fields. We used the stellar parameters for each star given in the TIC \citep{stassun2019}, which is based on data from Gaia Data Release 2 \citep{GaiaCollaboration2018}. The locations of the three ULTRASAT high cadence survey fields are given in Table \ref{tab:field_locations} and Figure \ref{fig:fields}.

We applied signal-to-noise cuts to the target list by estimating the signal-to-noise ratio (SNR) of a transiting planet with a radius of $1.5\ R_{\rm Jup}$ and an orbital period of 1 day around each star. We chose this planet as a limiting case for the largest reasonable SNR we would expect to see from a planet on any given star, allowing us to only include stars in the target list for which we can reasonably expect to detect planets. We used the 10th percentile Combined Differential Photometric Precision (CDPP; \citealt{Christiansen2012}) noise model from \cite{Kunimoto2022a} as the estimate of the noise for each star. We use this CDPP estimate and not the CDPP information from the actual {\tt tess-phomo} light curves we use for this search (see Section \ref{sec:tessphomo}) because, at the time this target list was created, we did not yet have enough {\tt tess-phomo} data to make accurate CDPP estimates. To determine whether this test planet would be detectable, we compare the estimated SNR of the transit signal to a detection threshold of SNR$=$7, which matches the detection threshold used by the {Kepler \it} pipeline. We cut all stars for which this estimated transit SNR is $<7$ from our sample, which left 646,936 stars. Finally, we cut any star that was either not observed by TESS or was never further than 9 pixels from the edge of a detector (and therefore was not fully captured), leaving us with a final sample of 639,736 stars.


\subsection{Light Curve Generation with tess-phomo} \label{sec:tessphomo}

To create light curves for each star in our target list, we used {\tt tess-phomo}\footnote{\url{https://github.com/robertfwilson/TESSphomo.git}} running on the Roman Research Nexus (RRN), a cloud-based data analysis environment developed and provided to the community by the Space Telescope Science Institute (STScI). {\tt tess-phomo} is user friendly, computationally reasonable, and optimized for faint stars by producing light curves with improved precision. Therefore, we use it to obtain FFI light curves for every star in our sample in every sector that star is observed. Wilson et al. (in prep.) provides a complete description of {\tt tess-phomo}; we provide a brief overview here. 

We started by cutting out a region of pixels in the TESS FFI around the desired target in a single observing sector - we chose a 25x25 pixel square cutout for all stars more than 13 pixels from the edge of a detector, for stars between 9 and 13 pixels we used a rectangular cutout truncated along one axis. This cutout size balanced computational complexity and execution speed. While TESS cutouts are available from an online service at the Barbara A. Mikulski Archive for Space Telescopes (MAST) at STScI \citep{Brasseur2019}, this implementation became impractical when dealing with large numbers of targets because each cutout generated its own file which led to an I/O bottleneck in our pipeline. Because we did not need to save the cutouts, we instead used \texttt{tesscube}. \texttt{Tesscube} efficiently extracts the cloud-based cutouts and keeps them in memory \citep{tesscube2025}. Each cutout took about 1--3 seconds to generate per sector.

{\tt tess-phomo} queries the TIC  for all stars brighter than Tmag=20.0 in the cutout. It then uses the mission delivered TESS Pixel Response Function (PRF) model to fit the target stars, as well as the sum of all background stars\footnote{\url{https://archive.stsci.edu/missions/tess/models/prf_fitsfiles/}}. {\tt tess-phomo} assumes that the relative flux of all background stars are constant, and uses those values to fit for the zeropoint-flux in the cutout. To account for PRFs that are centered outside our cutout but have wings within the cutout, we searched for all stars within a radius of 21.5 pixels of the center of the cutout (e.g., a minimum 7.5 pixel buffer). The full background model combines the background stars with a model for scattered light. {\tt tess-phomo} performs both aperture photometry and PRF photometry to extract the flux of the target star. For each star in our target list, we calculate light curves using both aperture and PRF photometry. There is currently no best practice for determining which of the two light curves is better to use. Therefore, for each target, we calculate the standard deviation of each light curve and use the one with the smaller standard deviation in our search. We note that the vast majority of our targets were searched using the PRF light curves, which is consistent with our expectation that the PRF photometry is generally better for faint targets \citep{Stetson1987}.


\subsection{Identifying Systematics-Driven Transit Signals}
\label{subsec:excess_tces}

Considering the large number of stars in our target list, we aimed to reduce the number of false alarms flagged in our transit search. For each TESS sector included in our sample, we determine the number of identified transit signals that fall on each time stamp. This method, called the Skye metric, was originally used in \cite{Thompson2018} when analyzing {\it Kepler} data and has been used subsequently for K2 and TESS. If every light curve in each sector contains only astrophysical signals (transits, eclipses, astrophysical noise, etc.), then we expect the number of transit signals should be equally distributed across time stamps (within some white-noise range). However, most systematic/instrumental effects will impact many light curves in a given sector at the same time, leading to a buildup of signals at specific time stamps.

\begin{figure*}
    \centering
    \includegraphics[width=0.9\linewidth]{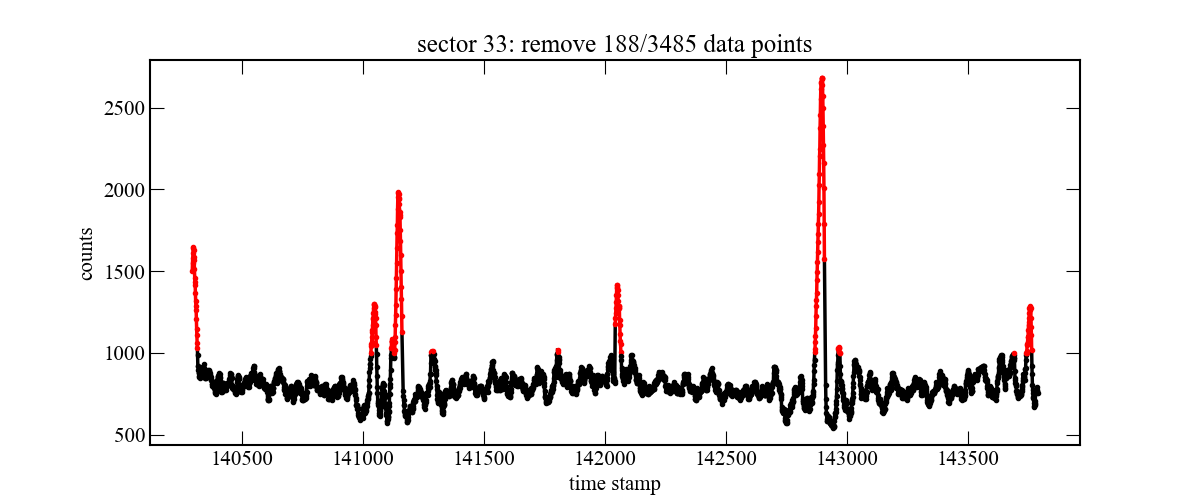}
    \caption{A small percentage of time stamps in a given sector generate large numbers of TCEs, indicating a systematic/instrumental effect at that cadence. This plot shows the number of times a given time stamp in sector 33 was flagged. The red points are time stamps that are masked before performing the full transit search. 188 of the 3485 time stamps are masked, or 5.4\% of the light curve.}
    \label{fig:excess_tce}
\end{figure*}

Out of our sample of 16 million light curves, we selected a random sample of $\sim10,000$ in each sector and ran a search using the transit least squares (TLS) algorithm on each light curve \citep{Hippke2019_tls}, obtained the most likely transit signal from each light curve, and determined the number of times each time stamp was flagged in a transit signal. Figure \ref{fig:excess_tce} shows the number of flagged events per time stamp in sector 33. For light curves with only white noise and randomly distributed transiting planets, the histogram should be flat. However, we see several peaks in the histogram indicating the build up of signals at specific time stamps. We chose to mask every time stamp for which the counts is $>3\sigma$ from the median value of the histogram for each sector, which we show in red in Figure \ref{fig:excess_tce}. We masked 5-15\% of all data points in a given sector, which is consistent with the number of data points masked in TESS data releases\footnote{\url{https://archive.stsci.edu/tess/tess_drn.html}}.

\subsection{Searching for Transits}
\label{subsec:tls}

We built a pipeline to search the {\tt tess-phomo} generated light curves for transits. First, we detrended the light curve to remove stellar variability and remaining instrument systematics using a time-windowed slider (window length of 0.8 days) utilizing Tukey's biweight method \citep{mosteller1977}, as implemented in the {\tt wotan} package \citep{hippke2019_wotan}, which is recommended specifically for use with the TLS algorithm we used for our transiting planet search (see below) \citep{hippke2019_wotan, Hippke2019_tls}. The removal of systematic trends is necessary for the transit search algorithm to detect the short-timescale and often low SNR transit signals.

\begin{figure*}
    \centering
    \includegraphics[width=0.9\linewidth]{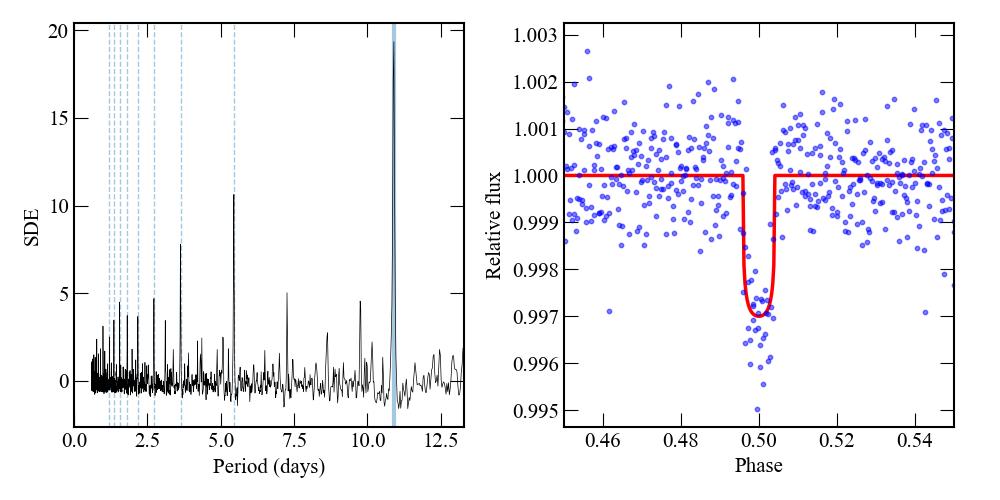}
    \caption{We were able to recover the known planet TOI-1266 b \citep{Demory2020} in the light curve from TESS sector 41. The left panel shows the TLS periodogram, with the period with the highest signal detection efficiency (SDE) shown in blue. The vertical dashed lines denote aliases of the detected period. The right panel shows the light curve phase folded on the period with the maximum SDE and the corresponding transit model in red.}
    \label{fig:tls}
\end{figure*}

After detrending the light curves, we searched for planets using the transit least squares (TLS) algorithm \citep{Hippke2019_tls}. TLS is similar to a traditional Box-Least Square (BLS) algorithm \citep{Kovacs2002}, but instead of using a rectangular box template to look for transit-like signals, it utilizes a template that includes more realistic, transit-like features, including limb-darkening effects and planetary ingress and egress. TLS uses the test statistic, the signal detection efficiency (SDE), which gives the statistical significance of a given transiting exoplanet orbital period compared to the mean significance of all other orbital periods. Figure \ref{fig:tls} shows the results of a TLS search of the light curve of TOI-1266 from TESS sector 41. This star hosts a known transiting planet (TOI-1266 b) with a 10.9 day orbital period \citep{Demory2020}. TLS is able to recover the known planet. Following \cite{Hippke2019_tls}, we set a detection threshold of SDE $=$ 7.

Since the ULTRASAT high-cadence survey fields are near the ecliptic poles, they have been observed in many TESS sectors. Depending on their exact location in the fields, stars in the target list have been observed in as few as 4 and as many as 46 sectors with TESS. The increased baseline allows us to observe more transits of a given planet and build up sufficient signal-to-noise to detect smaller planets. Therefore, combining all available light curves into a single TLS search would lead to the highest number of detected planets. However, this approach is very computationally expensive. The computational time of a TLS search is nonlinear with the light curve baseline. 

\begin{figure}
    \centering
    \includegraphics[width=\linewidth]{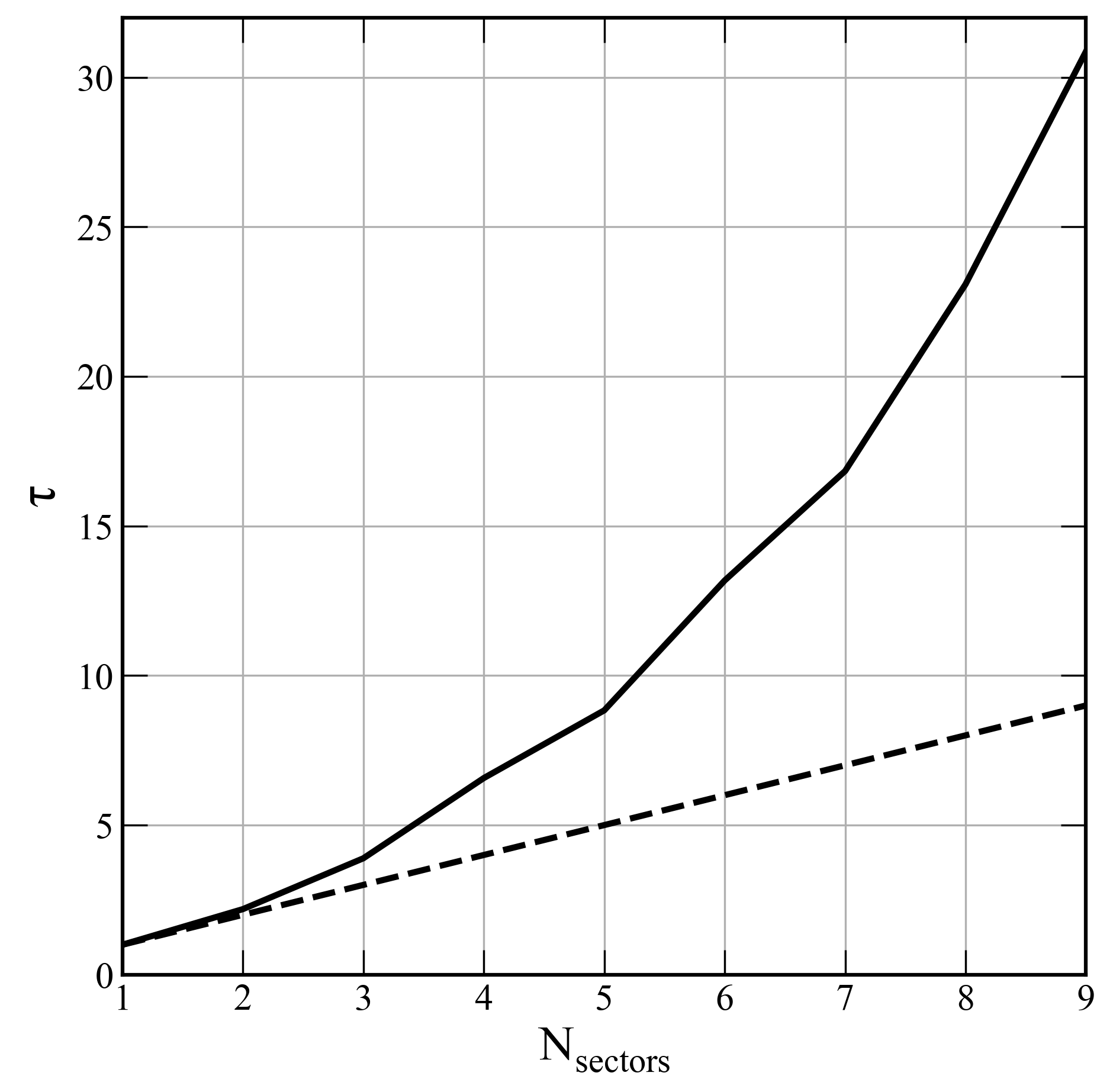}
    \caption{Stitching light curves into a single TLS search exponentially increases the computational time compared to running searches on each sector individually. This plot shows the runtime $\tau$ for TLS searches on light curves of TIC 149179763 normalized to the runtime for a TLS search of a single sector. The solid line shows the runtime after stitching the sectors into a single light curve, while the dashed line shows the runtime for individual TLS searches for each sector.}
    \label{fig:tls_runtime}
\end{figure}

To illustrate this, we examined light curves for TIC 149179763. This target was observed in 29 individual sectors, including all sectors from 31 to 39. We determined the runtime of a TLS search on the light curve from sector 31. We then stitched together sectors 31 and 32 and calculated the runtime of a TLS search. We repeated this process, adding another sector onto the light curve each time. The results are shown as the solid line in Figure \ref{fig:tls_runtime}, where $\tau$ is the computational time normalized to the time to run the search on just sector 31. We compare this to the computational time to run the TLS search on each sector individually, shown as the dashed line. Stitching multiple sectors into a single TLS search results in an exponentially increasing computational time compared to searching each sector individually.

The exponential increase in computational time shown in Figure \ref{fig:tls_runtime} primarily arises from the increased baseline of the light curve, which means that small shifts in the period of a transit model can significantly change the locations of the transits at the end of the light curve. Therefore, accurately determining the best period requires sampling a finer period grid, significantly increasing the computational cost. Running the TLS search on each sector separately is more computationally efficient, however it loses the advantage of using the additional sectors to build up the signal-to-noise to detect lower SNR planets. We chose a compromise between these two extremes by only stitching light curves together if they were observed in consecutive sectors, allowing for some increase in sensitivity to smaller planets while still breaking up the data to reduce the computational cost. It is also unlikely that planets that require the stacking of transits from all available data to detect in TESS would be detectable with ULTRASAT. We therefore do not expect to miss any planets detectable with ULTRASAT from this simplification. Additionally, we restricted the TLS search to only look for periods $<$12 days.

For each star in our sample, we took every detrended light curve and stitched together consecutive sectors of observation and leave non-consecutive sectors as separate light curve segments. For each light curve segment, we performed a TLS search and obtained the signal with the highest SDE. If that SDE$>$7, we masked the signal and performed another TLS search. This process was repeated five times or until there was no signal with SDE$>$7. Then, we compared the SDE$>$7 signals found in each light curve segment. If the same signal was identified in at least 20\% of individual light curve segments (or in 2 segments if there were less than five total segments), that signal was classified as a TCE.

Our strategy for identifying TCEs involved splitting the light curve into multiple segments and running the TLS search individually for computational efficiency. If a spurious signal was only present in one or a few sectors, it would not have a large enough signal-to-noise over the full observing baseline to produce a detectable signal. To remove these signals, we perform another TLS search by stitching together all available light curves for each TCE. For this search to remain reasonably computationally efficient, we greatly restricted the range of periods TLS searched over to only periods within 5\% of the TCE period. TCEs with SDE $>$ 7 in this search move on to the vetting process. This step also produced more accurate transit parameters (such as depth, duration, and period) that were used in vetting.

\subsection{Vetting}
\label{subsec:vetting}

The vetting process aims to distinguish whether a detected event is a real transiting planet, a false positive (other astrophysical signals), or a false alarm (signals unrelated to an astrophysical signal) \citep{Kostov2019}. We performed vetting to classify each TCE using the default thresholds and tests of {\tt LEO-vetter}, an automated vetting pipeline that computes vetting metrics and tests them against a series of pass-fail thresholds \citep{Kunimoto2025}. See \cite{Kunimoto2025} for full descriptions of each test, which include examining the transit shape and symmetry, testing whether the TCE is dominated by a single event, comparing the SNRs of each transit individually to ensure they are consistent, testing whether the transit duration is nonphysical, checking for differences in the odd versus even transit depths, searching for significant secondary eclipses, checking whether the estimated radius of the planet is physically reasonable, searching for a centroid offset indicative of contamination from a nearby eclipsing signal. For the planet radius test, {\tt LEO-vetter} adopts an upper limit on planet radius of 2 $R_{\rm Jup}$. During target selection we optimized for 1.5 $R_{\rm Jup}$ to maximize potential SNR, but here in vetting, we do not exclude planets between 1.5 and 2 because they do exist, but are rare. A TCE is promoted to a planet candidate if it passes all {\tt LEO-vetter} tests.

\subsection{Validation with TRICERATOPS}
\label{subsec:triceratops}

For each planet candidate identified in our vetting pipeline, we also ran {\tt TRICERATOPS}. {\tt TRICERATOPS} uses a Bayesian framework to test whether the data can be explained by a number of scenarios that produce transit-like signals, such as a true transiting planet, an eclipsing binary, a transit or eclipse around a background star, etc. by calculating a likelihood for each scenario \citep{Giacalone2020, Giacalone2021}. For each candidate, we take the stitched light curve with all available sectors, then phase fold the light curve on the candidate period. We then crop the light curve, keeping 2 times the transit duration on each side of the transit. Finally, we bin the data such that there are 100 points in the light curve, then calculated the likelihoods for each scenario. We combined these likelihoods into a total false positive probability (FPP) and nearby false positive probability (NFPP). Based on \cite{Giacalone2021}, we set the threshold for a validated planet to be FPP $< 0.015$ and NFPP $< 10^{-3}$. 

We note that some of the planet candidates we identify have significantly large FPP and NFPP ($\geq 0.90$). However, we choose not to reject high FPP candidates from our sample for multiple reasons. A large number of our candidates are giant planets, which are difficult to validate due to the fact that giant planets, brown dwarfs, and low mass stars are virtually indistinguishable based on radius alone \citep{Giacalone2021}. We also use stellar parameters from the TIC, which does not contain the most accurate available stellar parameters for each star. Finally, we did not include any additional information on each candidate when running {\tt TRICERATOPS}, such as contrast curves or ground-based light curves that would improve the calculation of the {\tt TRICERATOPS} false positive probabilities. While we do not exclude high FPP candidates, we do include the calculated FPP and NFPP for all candidates in our catalog as a reference for future studies.

\subsection{Estimating ULTRASAT SNR}\label{subsec:ultrasat_snr}

To determine which targets will be detectable with ULTRASAT, we estimated the SNR for each transiting planet candidate. To estimate the SNR, we used {\tt ULTRASAT-utils}\footnote{\url{https://github.com/mrtommyb/ULTRASAT-utils}}, a Python package with useful tools for ULTRASAT precursor science that includes a method to estimate the SNR of a transit. {\tt ULTRASAT-utils} takes the V-band magnitude of the star, as well as $T_{\rm eff}$ and $\log g$, and uses PHOENIX stellar models \citep{Husser2013} to predict the stellar brightness in the ULTRASAT bandpass. Then, it uses this brightness to estimate the noise of an ULTRASAT light curve, $\sigma_{\rm ULTRASAT}$. The SNR of a single transit in ULTRASAT is then given by:

\begin{equation}\label{eq:ultrasat-utils}
    {\rm SNR}_{1} = \frac{\rm depth}{\sigma_{\rm ULTRASAT}} \sqrt{\frac{\rm duration}{\rm cadence}}
\end{equation}
If a star is brighter than ULTRASAT's saturation limit, this calculation will return SNR$_1=0$. To estimate the total SNR, we calculate

\begin{equation}\label{eq:ultrasat-totalsnr}
    {\rm SNR}_{\rm U} = {\rm SNR}_{1} \times \sqrt{N}
\end{equation}
where $N$ is the total number of transits observed. To get $N$, we assumed a nominal three-year mission and three observing fields, with the telescope rotating between the fields every six months. Therefore, we assumed each field was observed for 1 year in total. The ULTRASAT high cadence survey will be conducted for $\sim 21$ hours per day (87.5\% of the day), with the rest of the time dedicated to a low cadence survey of a larger area and Target of Opportunity (ToO) observations \citep{Shvartzvald2024}. Therefore, we assumed each star will be observed for 87.5\% of a year, or $\sim$ 319 days. The number of transits observed of a given planet is 

\begin{equation}\label{eq:number of transits}
    N = \frac{319\ \rm{ days}}{P}
\end{equation}
where $P$ is the orbital period of the planet.

We set a detectability threshold at ${\rm SNR}_{\rm U} > 10$. This threshold gives a lower limit on the number of detectable planets, as this calculation assumes that the transit depth of each planet will be the same in TESS and ULTRASAT, which is not necessarily true given that we expect that some of these planets may have significantly increased transit depths in the UV as seen in some previous observations \citep[e.g.,][]{Salz2019, Sreejith2023, Cilley2026}. We also acknowledge that ${\rm SNR}_{\rm U} > 10$ is a relatively high threshold for detectability, especially for planets where the period is known. However, given the significant uncertainties in the ${\rm SNR}_{\rm U}$ calculation (detailed below), we chose to set a conservative detectability threshold, while acknowledging that some planets with ${\rm SNR}_{\rm U} < 10$ may be detectable. 

The estimation of the ULTRASAT SNR of a planet carries significant uncertainties. The primary source of uncertainty is in the PHOENIX stellar models used to predict the brightness of a star in the ULTRASAT bandpass. These stellar spectra models are much more accurate in the visible and infrared than in the UV, primarily because of the lack of large numbers of high-quality UV spectra of stars, which do exist for the visible and IR \citep{France2016}. Another source of uncertainty is the predicted ULTRASAT noise. While the specifications of the ULTRASAT instruments are known, they are currently being built and have not been robustly tested. Until that testing is complete, the noise properties of ULTRASAT will be uncertain. Additionally, the $\sqrt{N}$ term in Equation \ref{eq:ultrasat-totalsnr} is the ideal theoretical scaling of the SNR with $N$ for white noise. However, given the uncertain noise properties of ULTRASAT, it is unknown if this scaling will hold for real ULTRASAT light curves, or if SNR will increase as a lower power of $N$ due to correlated noise of some kind.

\section{Results} \label{sec:results}

\begin{deluxetable}{c|cc}
\tablecaption{Numbers of new and existing candidates \label{tab:numbers}}
\tablewidth{0pt}
\tablehead{
 & \colhead{All Planet Candidates} & \colhead{${\rm SNR_U} > 10$}
}
\startdata
New candidates & 181 & 30 \\
Previous TOIs & 296 & 83 \\
\hline
\textbf{Total:} & 477 & 113 \\
\enddata
\tablecomments{We report both the total number of candidates and the number that have SNR$_{\rm U}$ above our detectability threshold.}
\end{deluxetable}

\begin{figure*}
    \centering
    \includegraphics[width=0.95\linewidth]{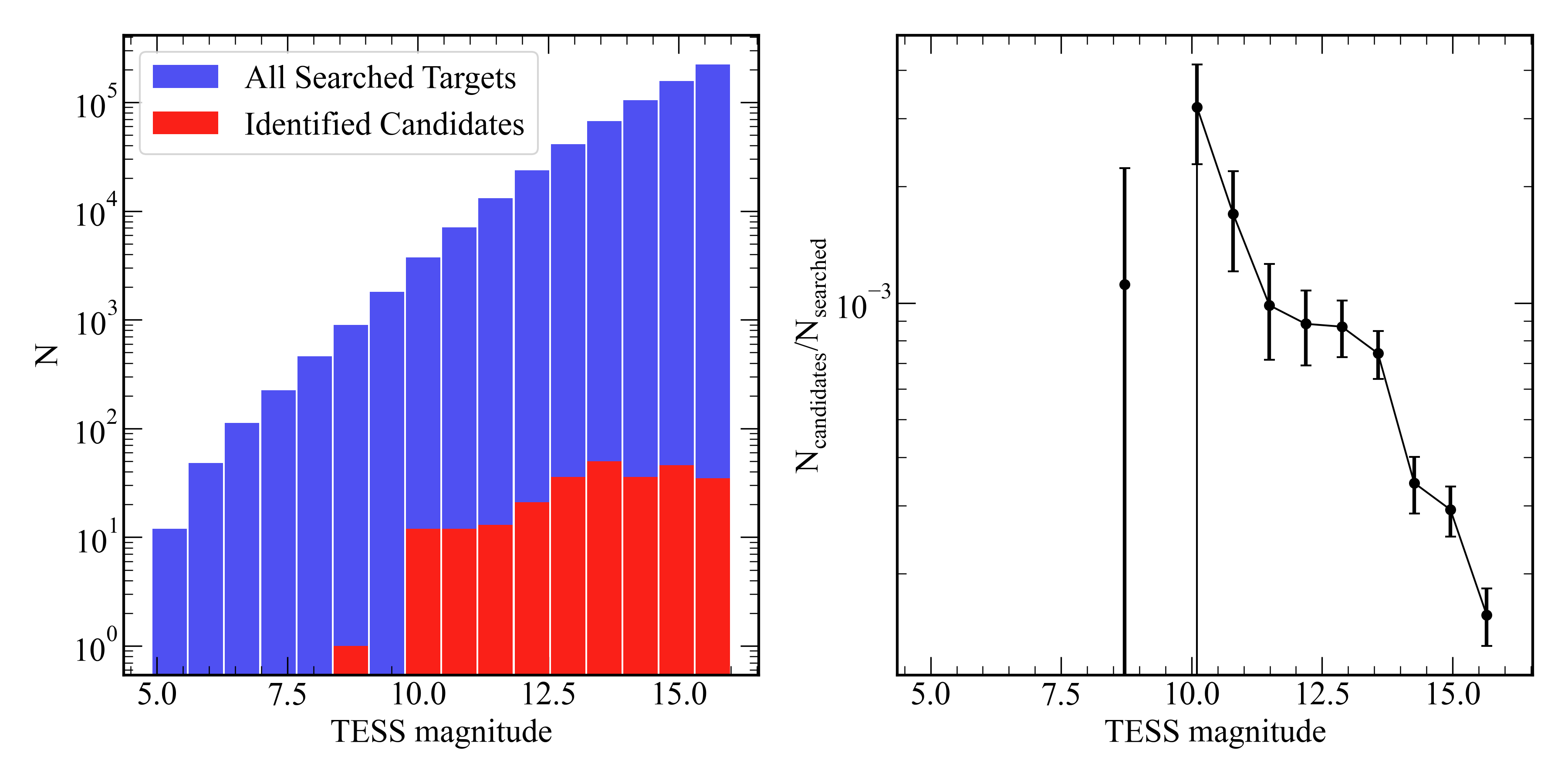}
    \caption{We detect a large number of planet candidates around faint stars due to the dramatic increase in the total number of stars with increasing magnitude. The left panel shows a histogram of the number of planet candidates detected as a function of TESS magnitude in red and the total number of stars searched in blue. The right panel shows the number of candidates divided by the total number of stars searched in each bin.}
    \label{fig:histogram}
\end{figure*}

Our initial search flagged 47,823 TCEs. Following the vetting process, we identify 46,447 as false alarms, 930 as false positives (FP), 184 as nearby false positives (NFP), and 262 as planet candidates (PC). Of the 262 identified as planet candidates, 181 are new to this work. Catalogs of these results are available as machine-readable tables accompanying this paper and are detailed in Appendix \ref{appx:tables}.

For all targets, both previously identified and new from this work, we calculate the ULTRASAT SNR using Eq. \ref{eq:ultrasat-totalsnr}. Table \ref{tab:numbers} shows the number of new and existing planet candidates that have SNR higher than the threshold. In total, 113 candidates will be detectable with ULTRASAT, including 30 new candidates and 83 previously known TOIs.

Figure \ref{fig:histogram} shows the distribution of identified planet candidates as a function of TESS magnitude. The right panel shows the number of identified candidates divided by the total number of targets searched in that bin. We find a generally decreasing event rate (i.e., candidates per star) as we go to fainter stars, which is consistent with our expectations. Despite the decreasing event rate, the total number of stars searched also increases, meaning we still detect large number of planet candidates around faint stars.

\begin{figure}
    \centering
    \includegraphics[width=\linewidth]{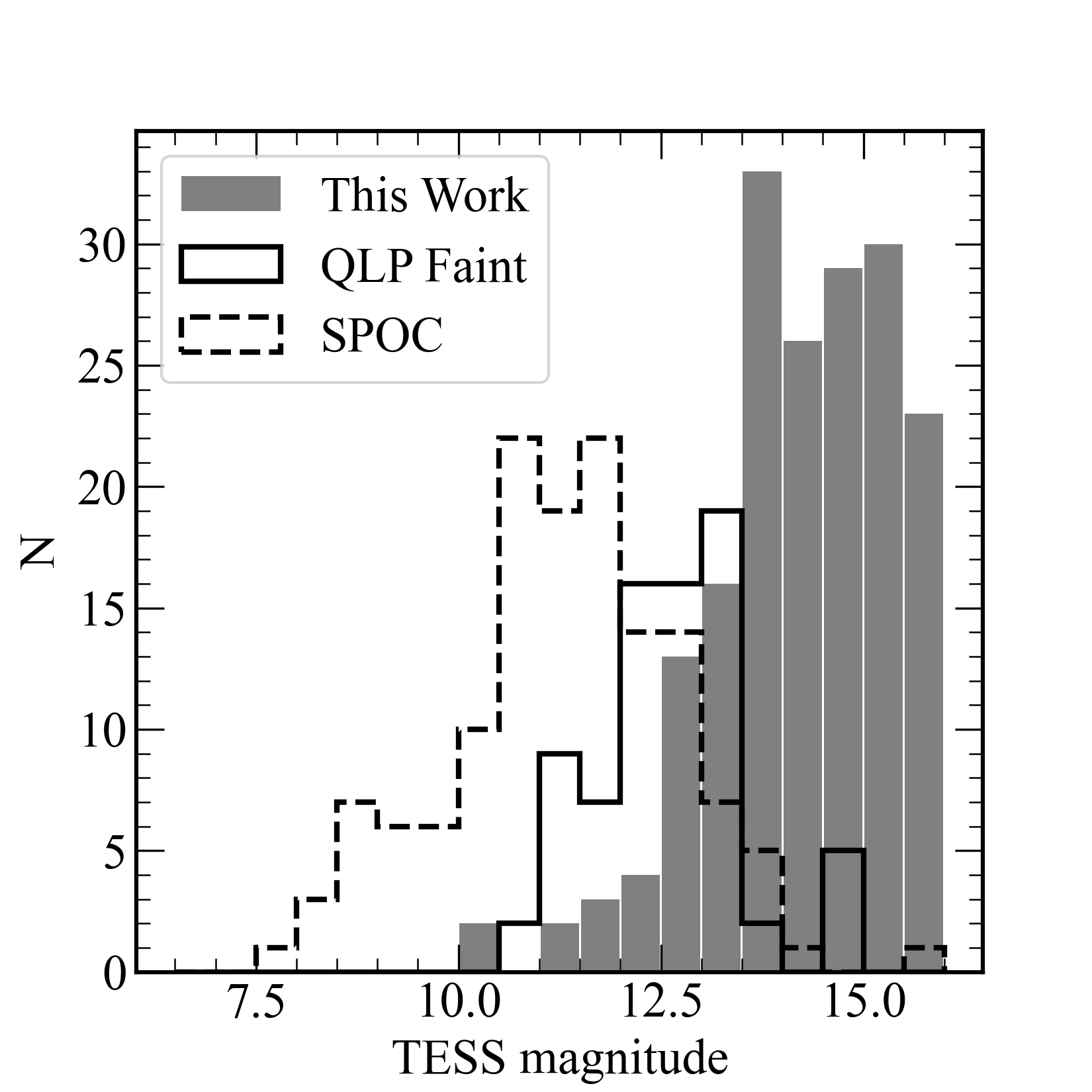}
    \caption{This work finds candidates around fainter stars than previous searches. This histogram compares the new sample of candidates from this work in the filled-in bars with the sample of candidates from the QLP faint star search in the solid line, and the sample of candidates found from SPOC in the dashed line. This figure includes only planets from SPOC and QLP that are in the ULTRASAT fields.}
    \label{fig:toi_comparison}
\end{figure}

We show a comparison between the new candidates found in this work with existing candidates on the TOI list in Figure \ref{fig:toi_comparison}. The previously identified TOIs are in the empty bars, and the new candidates from this work are in the filled-in bars. There is a steep drop in known TOIs at magnitude 13.5, the magnitude limit for the faint star search with QLP. We see that our new candidates are preferentially around fainter stars than the previous TOIs, including a significant number fainter than the 13.5 threshold. 

\begin{figure}
    \centering
    \includegraphics[width=\linewidth]{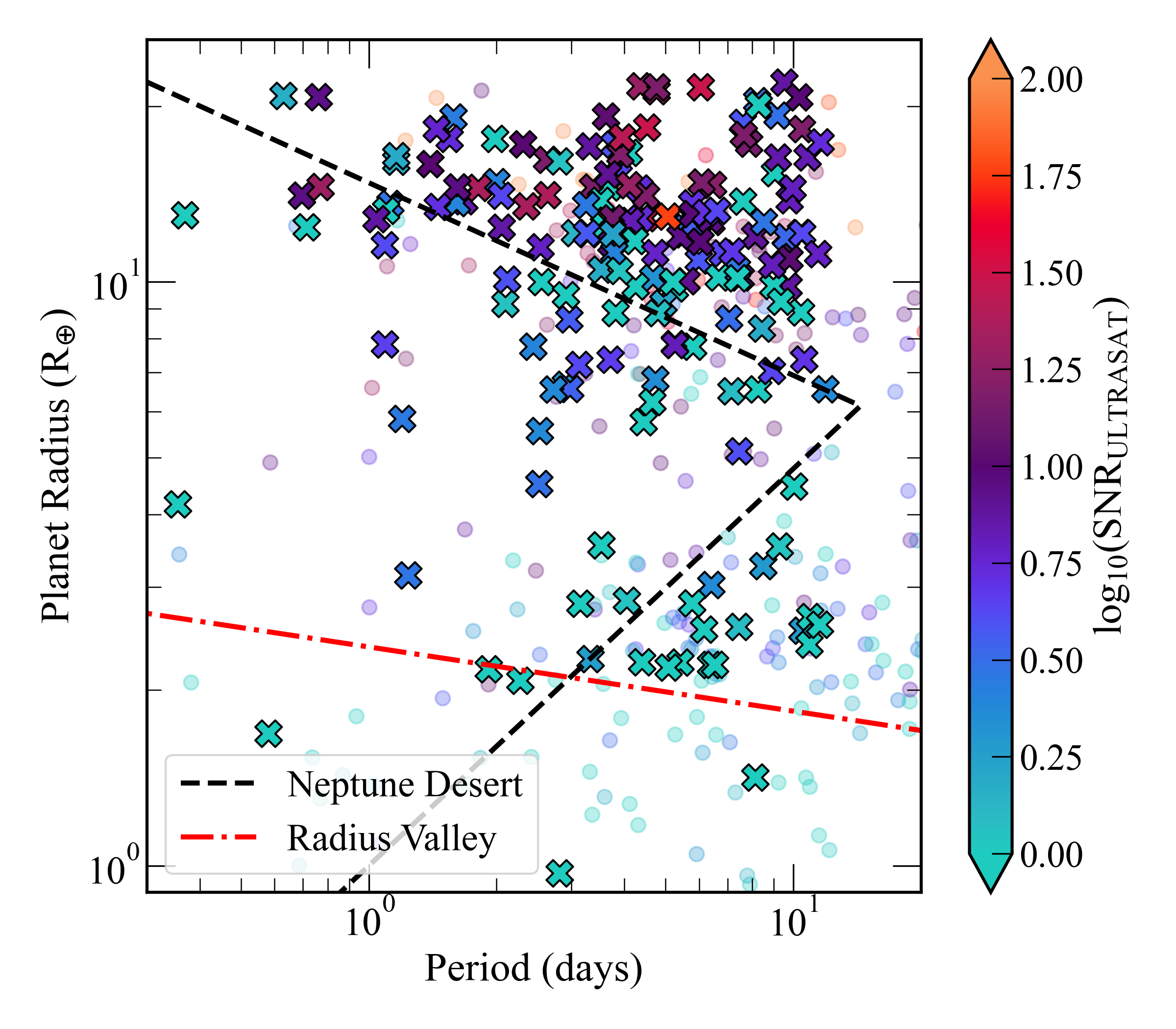}
    \caption{This work identifies many planet candidates in the Hot Neptune Desert and near the Radius Valley. New candidates from this work are plotted as crosses and previously identified TOIs are shown as transparent circles. Each point is colored by the ULTRASAT signal to noise (in log scale). We also plot the boundaries of the Hot Neptune Desert in black from \cite{Mazeh2016} and the location of the Radius Valley in red from \cite{Petigura2022}.}
    \label{fig:radius_valley}
\end{figure}

Figure \ref{fig:radius_valley} shows the new candidates from this work plotted with the previously known candidates in the ULTRASAT field as a function of planet radius and orbital period. Each data point is colored by the estimated ULTRASAT SNR as calculated in Section \ref{subsec:ultrasat_snr}. We find that most candidates detectable with ULTRASAT are Hot Jupiters. Figure \ref{fig:radius_valley} also shows the locations of the Hot Neptune Desert and the Radius Valley. We identify a number of candidates in the Hot Neptune Desert, a small number of which may be detectable with ULTRASAT under ideal conditions. This search also detects several sub-Neptune-sized and smaller candidates near the Radius Valley. While a TESS transit search of faint stars is sensitive to a small number of these candidates, ULTRASAT will likely not have the sensitivity necessary to detect them.

\begin{figure}
    \centering
    \includegraphics[width=\linewidth]{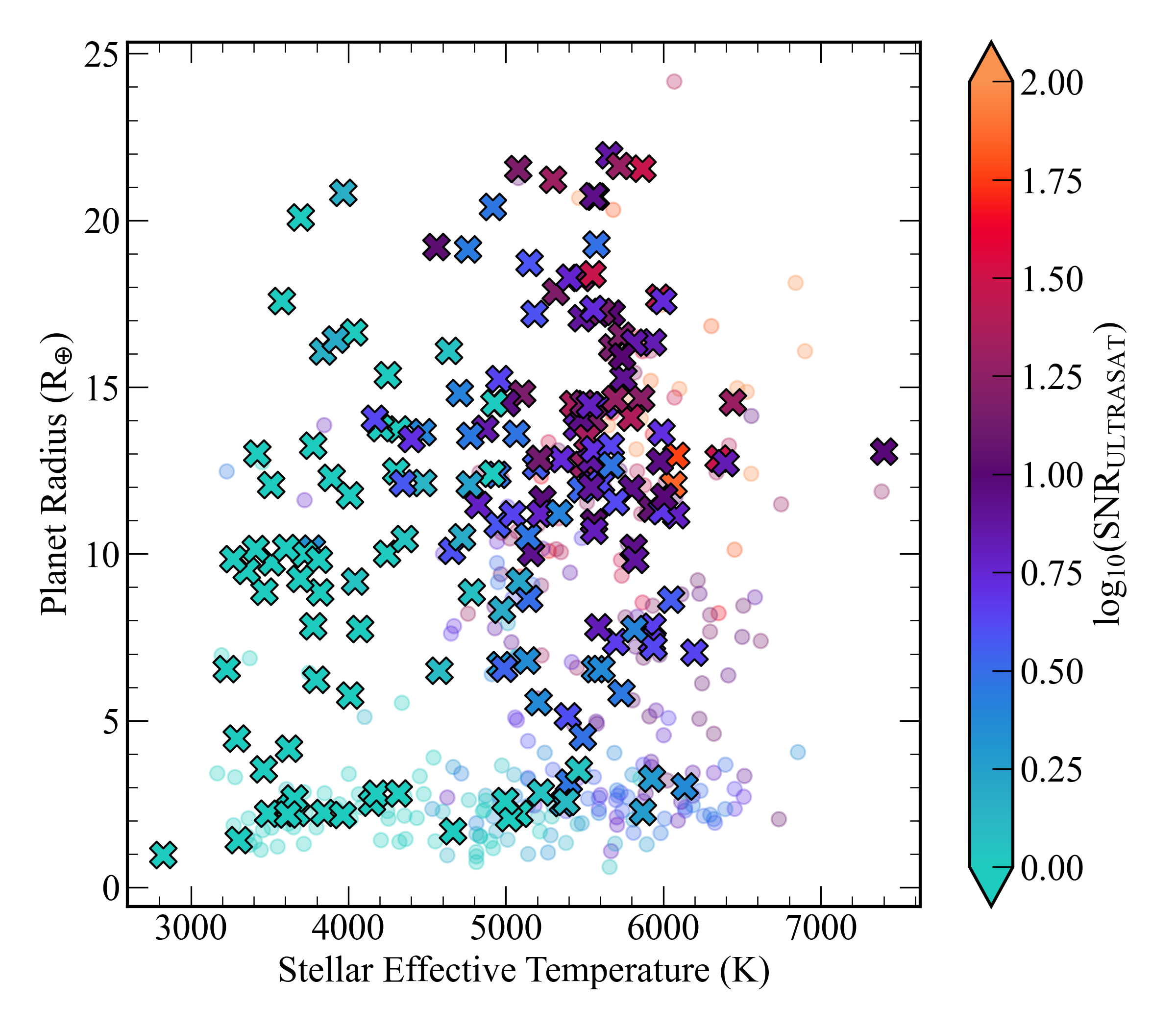}
    \caption{ULTRASAT will be more sensitive to planets around hotter stars. This plot shows planet candidates in the ULTRASAT fields as a function of planet radius and stellar effective temperature. The points are colored the same as Figure \ref{fig:radius_valley} showing the ULTRASAT signal to noise. }
    \label{fig:rp_v_teff}
\end{figure}

Figure \ref{fig:rp_v_teff} is similar to Figure \ref{fig:radius_valley} but shown as a function of planet radius and stellar effective temperature. We see a trend that planet candidates around hotter stars have higher ${\rm SNR}_{\rm U}$. Since hotter stars have SEDs that peak at bluer wavelengths, they will have a higher amount of emission in the UV compared to cooler stars. They will appear brighter in the ULTRASAT bandpass and thus have planet candidates that are more detectable. 

\section{Discussion}\label{sec:discussion}

\subsection{Validated Planets and Other Interesting Candidates}\label{sec:validated}

Following \cite{Giacalone2021}, a planet candidate is considered validated if it has FPP $< 0.015$ and NFPP $< 10^{-3}$. We ran {\tt TRICERATOPS} on all identified planet candidates, and found that 10 candidates passed the validation thresholds. Of these, 7 have $R_p>8R_{\oplus}$. Statistical validation with only a light curve is notoriously unreliable for giant planets, given that gas giant planets, brown dwarfs, and the lowest mass stars all have approximately the same radii. As in \cite{Giacalone2021}, we therefore do not consider any planet candidate having $R_p>8R_{\oplus}$ as validated, leaving 3 remaining (TIC 55401933.01, TIC 220432308.01, TIC 219893931.01).

\begin{figure*}
    \centering
    \includegraphics[width=0.95\linewidth]{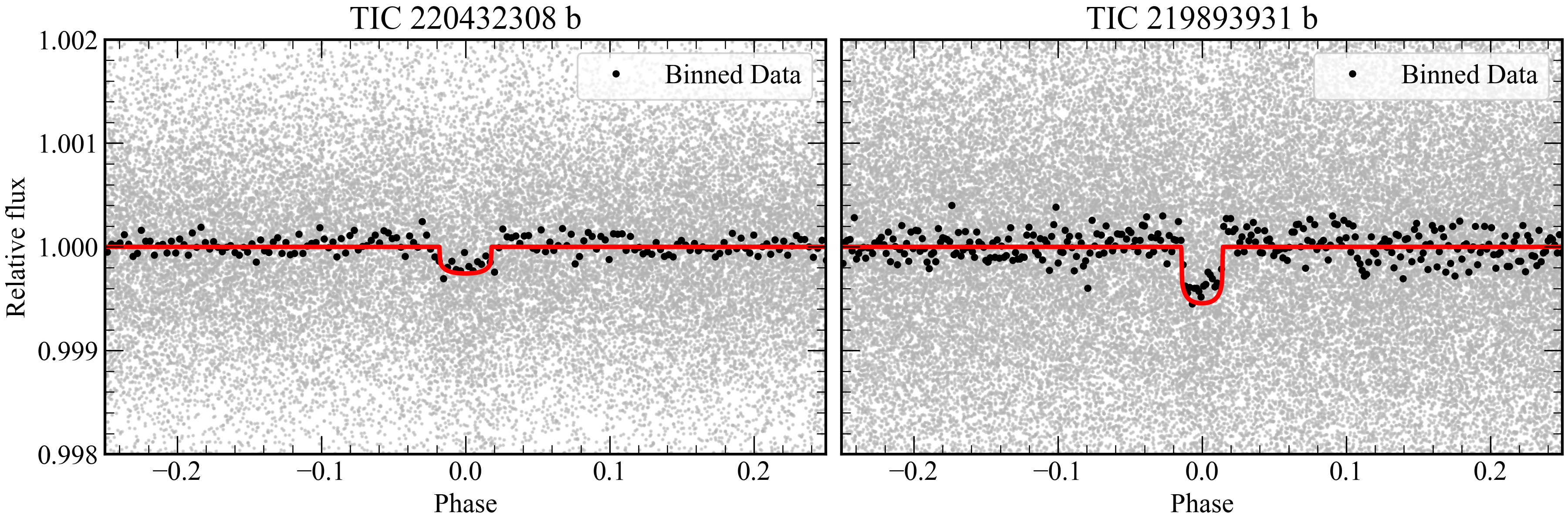}
    \caption{Phase folded light curves for TIC 220432308 b and TIC 219893931 b. The gray points are the full detrended light curve and the black points are binned data. The red curve is the transit model from {\tt LEO-vetter}. The parameters for each planet are found in Table \ref{tab:validated}.}
    \label{fig:validated_phase_fold}
\end{figure*}

For each of the remaining candidates, we ran it through validation with {\tt TRICERATOPS} 20 times to ensure the original FPP calculation is not an outlier, given that the FPP is expected to have intrinsic scatter in the calculation \citep{Giacalone2021}. We found that TIC 55401933.01, which had an initial FPP below the threshold (FPP$=0.0130$), had a mean FPP $=0.0534\pm0.0008$, which does not meet the validation threshold. We therefore do not consider that candidate validated. The two remaining candidates had a mean FPP below the threshold and are considered statistically validated, and therefore bona fide planets. We henceforth refer to them as TIC 220432308 b and TIC 219893931 b. Table \ref{tab:validated} shows the properties of each of the validated planets, and Figure \ref{fig:validated_phase_fold} shows the phase folded light curves for each planet. We note that both validated planets have SNR$_{\rm U}<2.5$, meaning that neither planet is likely detectable with ULTRASAT.

\begin{deluxetable}{c|cc}
\tablecaption{Validated Planets \label{tab:validated}}
\tablewidth{0pt}
\tablehead{
 & \colhead{TIC 220432308 b} & \colhead{TIC 219893931 b}
}
\startdata
TESS mag & 11.2562 & 11.6498 \\
RA (deg) & 73.251756 & 262.561894 \\
Dec (deg) & -59.392691 & 71.842937 \\
Period (days) & $3.330099\pm0.000008$ & $6.410510\pm0.000014$ \\
T$_0$ (BJD) & $1328.562 \pm 0.002$ & $1686.482\pm0.002$ \\
Depth (ppm) & 182.49 & 410.26 \\
Duration (day) & $0.098\pm0.229$ & $0.101\pm0.172$ \\
$R_p\ (R_{\oplus})$ & $2.26\pm0.53$ & $3.03\pm0.35$ \\
$a$ (AU) & $0.0445\pm0.0018$ & $0.0710\pm0.0033$ \\
$S_{\rm eff}$ ($S_\oplus$) & $1048\pm156$ & $410\pm65$ \\
$R_*\ (R_{\odot})$ & $1.40\pm0.07$ & $1.28\pm0.06$ \\
$T_{\rm eff}$ (K) & $5870\pm118$ & $6136\pm133$ \\
FPP & $(8.2\pm0.3)\times10^{-3}$ & $(7.6\pm1.1)\times10^{-3}$ \\
NFPP & $(8.9\pm0.1)\times10^{-4}$ & $(9.3\pm1.7)\times10^{-6}$ \\
SNR$_{\rm U}$ & 1.85 & 2.37 \\
\enddata
\tablecomments{The FPP and NFPP listed for each planet are the mean values after 20 runs of {\tt TRICERATOPS}.}
\end{deluxetable}

Among our newly identified planet candidates, we also note some of the most interesting candidates that are worthy of further study.

TIC 198079261.01 is the smallest planet candidate detected in our survey. We calculate a planet radius of 0.97 $R_{\oplus}$ and an orbital period of 2.82 days. After running {\tt TRICERATOPS}, we find FPP$=0.127$ and NFPP$=0.023$. Stellar parameters from the TIC indicate that the host star has a radius of 0.123 $R_{\odot}$, placing it among the smallest stars for which we have detected transiting exoplanets (TRAPPIST-1 has $R_s=0.119\ R_{\odot}$; \citealt{Gillon2017,Agol2021}). TIC 198079261.01 also has TSM$=$12.3, indicating that it is amenable to atmospheric characterization through transmission spectroscopy with JWST ($J=12.9$). 

TIC 199608547.01 straddles the Radius Valley with $R_p=2.17\ R_{\oplus}$ and is on a short orbital period of 1.92 days, making it a valuable system for assessing planet formation and evolution. The host star is an early-type M dwarf star with $R_s=0.56\ R_{\odot}$. It has {\tt TRICERATOPS} FPP$=0.068$ and NFPP$=0.024$. 

TIC 358337224.01 is a giant planet candidate ($R_p = 1.8\ R_{\rm Jup}$) orbiting an M dwarf ($R_s = 0.58\ R_{\odot}$) with an orbital period of 8.29 days. The {\tt TRICERATOPS} probabilities are FPP$=0.221$ and NFPP$=0.014$. Planet formation models struggle to predict how planets of this size can form around such small stars \citep[e.g.,][]{Endl2006}. We caution that this candidate has a relatively high false positive probability and urge further confirmation that the observed transit is planetary in nature.

TIC 219009050.01 is one of several new candidates detected in the hot Neptune Desert \citep[e.g.,][]{Mazeh2016}. We calculate a radius of 7.82 $R_{\oplus}$ and an orbital period of 1.09 days. The {\tt TRICERATOPS} probabilities are FPP$=0.071$ and NFPP$=0.021$.

\subsection{Comparison to the T16 Search}\label{sec:t16}

Recently, the T16 project published results from a transit search of light curves from TESS Cycle 1 down to TESS magnitude 16 where they identified more than 10,000 planet candidates \citep{Hartman2025, Roth2026}. Since TESS Cycle 1 overlaps with the S1 ULTRASAT field, we briefly compare their results with ours. There are 263 T16 candidates in the S1 ULTRASAT field. Of these, 23 match candidates from this work. We determine that our search is much more conservative than the T16 search when identifying planet candidates. To demonstrate this, we compare the T16 candidates to all TCEs identified in our search. We find that 206 of our TCEs match a T16 candidate. Most of these are classified as either false alarms (89) or false positives (83) by {\tt LEO-vetter}. An additional 11 were classified as nearby false positives.

\section{Conclusions} \label{sec:conclusions}

In this work, we present a catalog of transiting exoplanets in the ULTRASAT high cadence survey fields. We detect 181 new planet candidates and statistically validate 2 of those candidates. This work demonstrates the power of TESS to detect exoplanets around faint stars. We show that while the candidate event rate of a transit search decreases for fainter stars, the increasing number of stars still allows for a large number of planet candidates to be detected. This work also nearly doubles the number of planet candidates detectable with ULTRASAT. This catalog of well-vetted transiting planet candidates will lead to the largest sample of planets with observations in the UV.

This work also demonstrates the power of pushing TESS transit searches to fainter stars. Performing such a search on the entire TESS dataset would produce a large number of new planet candidates, which we estimated using a brief, back-of-the-envelope calculation. The 3 ULTRASAT high cadence survey fields cover $\approx$612 square degrees, which is $\sim$1.5\% of the entire sky. Assuming the detection efficiency is the same across the entire sky, we estimate that the search would detect $\sim$12,000 planet candidates, which is significantly larger than the total number of planet candidates identified by TESS to date. This sample would probe a larger volume of the Galaxy than the current TESS sample and significantly increase the number of M dwarfs with observed planet candidates. Any such search would require significant optimization of the transit search and vetting algorithms and significant computational resources, but has the potential to significantly increase the science yield of the TESS mission. 

There are several possible extensions of this work, including searching for longer period planets and additional planets within these systems. We restricted our search to only look for planets with periods $<$12 days. However, since every star in our sample has been observed multiple times with TESS, the data exist to detect longer-period planets. Additionally, our pipeline is not currently built to identify systems with multiple planets. These systems are interesting as they provide important clues toward understanding the formation and evolution of planetary systems.

\begin{acknowledgments}

We would like to thank the referee for providing comments and suggestions that improved the paper.
The material is based upon work supported by NASA under award number 80GSFC24M0006.
This work was supported by the Ultraviolet Transient Astronomy Satellite Participating Scientists Program. We gratefully acknowledge the support of the ULTRASAT Science Team.
This work made use of the Roman Research Nexus (RRN), a JupyterHub service provided by the Space Telescope Science Institute (STScI). 
This research has made use of the NASA Exoplanet Archive, which is operated by the California Institute of Technology, under contract with the National Aeronautics and Space Administration under the Exoplanet Exploration Program. 
This paper includes data collected with the TESS mission, obtained from the MAST data archive at the Space Telescope Science Institute (STScI). Funding for US Institutions for the TESS mission is provided by the NASA Explorer Program. STScI is operated by the Association of Universities for Research in Astronomy, Inc., under NASA contract NAS 5–26555.

\end{acknowledgments}

\facilities{TESS \citep{Ricker2015}}

\software{
astropy \citep{astropy13,astropy18, astropy22}, 
batman \citep{Kreidberg2015},
Jupyter \citep{jupyter}, 
LEO-vetter \citep{Kunimoto2025},
Lightkurve \citep{lightkurve}, 
Matplotlib \citep{matplotlib},
NumPy \citep{numpy}, 
Pandas \citep{pandas}, 
scipy \citep{scipy},
tess-phomo (\url{https://github.com/robertfwilson/TESSphomo.git}),
transitleastsquares \citep{Hippke2019_tls},
triceratops \citep{Giacalone2020, Giacalone2021},
wotan \citep{hippke2019_wotan},
}

\appendix
\section{Tables of Transit Search Results}\label{appx:tables}

\begin{deluxetable}{lllll}
\tablecaption{New Planet Candidates\label{tab:new_cands}}
\tablewidth{0pt}
\tablehead{
\colhead{Column} & \colhead{Units} & \colhead{Label} & \colhead{Description}
}
\startdata
1  & \nodata & TICID       & TIC Identifier \\
2  & mag     & TESSmag     & TESS magnitude \\
3  & deg     & ra          & Right Ascension (J2000) \\
4  & deg     & dec         & Declination (J2000) \\
5  & days    & period      & Orbital period ($P$)\\
6  & days    & e\_period   & Uncertainty in $P$ \\
7  & BTJD     & T0          & Transit epoch ($T_0$) \\
8  & days    & e\_T0       & Uncertainty in $T_0$ \\
9  & ppm     & depth       & Transit depth \\
10 & days    & duration    & Transit duration ($t_{dur}$) \\
11 & days    & e\_duration & Uncertainty in $t_{dur}$ \\
12 & $R_\oplus$ & Rp       & Planet radius ($R_p$) \\
13 & $R_\oplus$ & e\_Rp    & Uncertainty in $R_p$ \\
14 & AU      & a           & Semimajor axis ($a$) \\
15 & AU      & e\_a        & Uncertainty in $a$ \\
16 & $S_\oplus$      & Seff           & Insolation ($S_{\rm eff}$) \\
17 & $S_\oplus$      & e\_Seff        & Uncertainty in $S_{\rm eff}$ \\
18 & $R_\sun$   & Rstar    & Stellar radius ($R_*$) \\
19 & $R_\sun$   & e\_Rstar & Uncertainty in $R_*$ \\
20 & K       & Teff        & Effective temperature ($T_{\rm eff}$) \\
21 & K       & e\_Teff     & Uncertainty in $T_{\rm eff}$ \\
22 & \nodata & FPP         & {\tt TRICERATOPS} FPP \\
23 & \nodata & NFPP        & {\tt TRICERATOPS} NFPP \\
24 & \nodata & SNR\_U      & Expected ULTRASAT SNR \\
\enddata
\tablecomments{This table only includes planet candidates that are new to this work. 
Table \ref{tab:new_cands} is published in its entirety in machine-readable format in the electronic edition of the journal and at \url{https://zenodo.org/records/23504163}. A portion is shown here for guidance regarding its form and content.}
\end{deluxetable}

Tables containing the results of of our transit search are available in machine-readable format accompanying this paper. Table \ref{tab:new_cands} lists and describes the parameters available in the catalog of new planet candidates in this work. We note that this table also includes the two validated planets. 

Our vetting pipeline classified 1114 TCEs as either false positives or nearby false positives, the vast majority of which are likely to be eclipsing binaries. While these objects are outside the scope of this work, they will also be interesting systems to observe with ULTRASAT. We therefore include a table containing every TCE from our search classified as either FP, NFP, or PC. Table \ref{tab:all_tces} lists and describes the parameters available in that table. We note that this table has not been cross-matched with the TOI list and therefore includes some events that correspond to previously identified TOIs.

\begin{deluxetable}{lllll}
\tablecaption{All TCEs Classified as FP, NFP, or PC\label{tab:all_tces}}
\tablehead{
\colhead{Column} & \colhead{Units} & \colhead{Label} & \colhead{Description}
}
\startdata
1  & \nodata & TICID       & TIC Identifier \\
2  & mag     & TESSmag     & TESS magnitude \\
3  & deg     & RA          & Right Ascension (J2000) \\
4  & deg     & Dec         & Declination (J2000) \\
5  & days    & period      & Orbital period ($P$)\\
6  & days    & e\_period   & Uncertainty in $P$ \\
7  & BTJD     & T0          & Transit epoch ($T_0$) \\
8  & days    & e\_T0       & Uncertainty in $T_0$ \\
9  & ppm     & depth       & Transit depth \\
10 & days    & duration    & Transit duration ($t_{dur}$) \\
11 & days    & e\_duration & Uncertainty in $t_{dur}$ \\
12 & $R_\oplus$ & Rp       & Planet radius ($R_p$) \\
13 & $R_\oplus$ & e\_Rp    & Uncertainty in $R_p$ \\
14 & AU      & a           & Semimajor axis ($a$) \\
15 & AU      & e\_a        & Uncertainty in $a$ \\
16 & $S_\oplus$      & Seff           & Insolation ($S_{\rm eff}$) \\
17 & $S_\oplus$      & e\_Seff        & Uncertainty in $S_{\rm eff}$ \\
18 & $R_\sun$   & Rstar    & Stellar radius ($R_*$) \\
19 & $R_\sun$   & e\_Rstar & Uncertainty in $R_*$ \\
20 & K       & Teff        & Effective temperature ($T_{\rm eff}$) \\
21 & K       & e\_Teff     & Uncertainty in $T_{\rm eff}$ \\
22 & \nodata & FPP         & {\tt TRICERATOPS} FPP \\
23 & \nodata & NFPP        & {\tt TRICERATOPS} NFPP \\
24 & \nodata & Disp        & Flag indicating FP, NFP, PC \\
\enddata
\tablecomments{Table \ref{tab:all_tces} is published in its entirety in machine-readable format in the electronic edition of the journal and at \url{https://zenodo.org/records/23504163}. A portion is shown here for guidance regarding its form and content.}
\end{deluxetable}

\bibliography{references}{}
\bibliographystyle{aasjournalv7}

\end{document}